%% file: main.tex
\documentclass[12pt]{article}
\usepackage{graphicx} 
\usepackage{amsmath}
\usepackage{authblk}
\usepackage{amssymb}
\usepackage{amsfonts}
\usepackage{natbib}
\usepackage{float}
\usepackage{xcolor}
\usepackage[margin=1in]{geometry}
\usepackage{rotating}
\newtheorem{lemma}{Lemma}

\begin{document}

\title{Hazard-Based Targeted Maximum Likelihood
Estimation for Survival in Resampling Designs}
\author[1]{Kirsten E. Landsiedel}
\author[1]{Rachael V. Phillips}
\author[1]{Maya L. Petersen}
\author[1]{Mark J. van der Laan}
\affil[1]{Division of Biostatistics, University of California, Berkeley}

\maketitle

\begin{abstract}

Survival is a key metric for evaluating current standards of care for people living with HIV. In resource-limited settings, high rates of loss to follow-up (LTFU) often result in underestimation of mortality when only observed deaths are considered. Resampling (also known as double sampling or two-stage sampling), which tracks a subset of LTFU patients to ascertain their outcomes, mitigates bias and improves survival estimates. However, common estimators for survival in resampling designs, such as weighted Kaplan-Meier (KM), fail to leverage covariate information collected during repeated clinic visits, even though this information is highly predictive of survival.
We propose a novel Targeted Maximum Likelihood Estimator (TMLE) for survival in resampling designs, which addresses these limitations by leveraging baseline and longitudinal covariates to achieve greater efficiency. Our TMLE is a plug-in estimator that respects survival curve monotonicity and is robust to misspecification of the initial model for the conditional hazard of death, guaranteeing consistency of our estimator due to known resampling probabilities.
We present: (1) a fully efficient TMLE for data from resampling studies with fixed follow-up time for all participants and (2) an inverse probability of censoring weighted (IPCW) TMLE that accounts for varied follow-up times by stratifying on patients with sufficient follow-up to evaluate survival. This IPCW-TMLE can be made highly efficient through nonparametric or targeted estimation of the follow-up censoring mechanism. 
In simulations mimicking real-world data, our TMLE reduced variance by as much as 55\% compared with the commonly used weighted Kaplan-Meier estimator while preserving nominal 95\% confidence interval coverage. These findings demonstrate the potential of our TMLE to improve survival estimation in resampling designs, offering a robust and resource-efficient framework for HIV research and other applications. 


\end{abstract}

\newpage

\section{Introduction} \label{sec:intro}

During HIV treatment, patients attend regular clinic visits to monitor treatment efficacy. However, estimating mortality from passively collected clinical and death reporting data is challenging due to high rates of loss to follow-up (LTFU) and administrative censoring (in resource-limited settings, rates of LTFU commonly range from 15\% to 50\% \citep{geng2015estimation, geng2012causal, yiannoutsos2008sampling}). Prior research has shown that patients who become LTFU in this setting often experience higher mortality rates and more severe illness, leading to a substantial underestimation of overall mortality when relying solely on observed data \citep{holmes2018estimated}. Methods for estimation within this context are limited given that LTFU cannot directly be treated as right censoring given that death may itself be the cause of LTFU \citep{geng2012causal}. This violates the typical assumptions of independent or conditionally independent censoring with respect to the outcome.

To address this issue, resampling designs introduce a second stage of data collection in which a subset of LTFU patients (rather than all, due to resource limitations) are systematically tracked to determine their mortality status. Incorporating resampled data helps reduce bias in survival estimates derived solely from observed data. One study found that incorporating resampled data increased the estimated mortality rate from 1.9\% to 7.0\%, which speaks to the extent to which mortality may be under-reported when using clinical data alone \citep{holmes2018estimated}. While existing methods can produce unbiased estimates of survival in resampling designs, a key methodological challenge lies in constructing efficient estimators which fully leverage baseline and time-dependent covariate information, a gap which we aim to fill.

\begin{figure}[htbp]
    \centering
    \includegraphics[width=\linewidth]{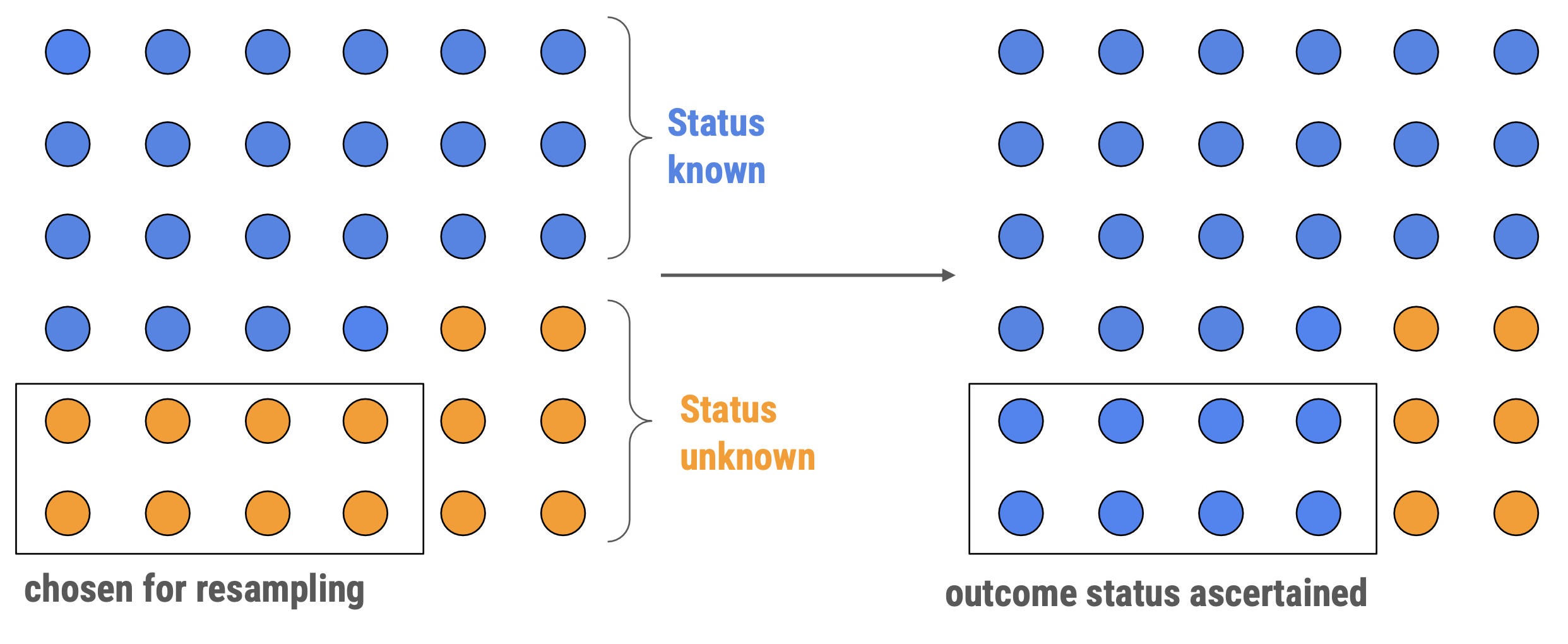} 
    \caption{Depiction of resampling process. Each dot represents one or more participants in the study. Blue dots represent participants with known outcomes. Orange dots represent participants with unknown outcomes.}
    \label{fig:your_label} 
\end{figure}

To our knowledge, no semiparametric efficient estimators exist that fully incorporate longitudinal covariate information in survival estimation under resampling designs. The prevailing estimators (weighted KM, weighted hazard averages, deductive estimators) either discard available covariate data, rely on strong parametric assumptions, or preclude closed-form efficiency analysis. To address this gap, we propose two estimators for survival in resampling studies: (1) a semiparametric efficient TMLE for fixed follow-up designs, and (2) an IPCW-TMLE that accommodates varied follow-up times. Both estimators rely on analytic influence curve derivation and flexibly incorporate time-varying covariates through Super Learner, yielding estimators that are doubly robust, more efficient, and tailored to the complex data structures arising in HIV care programs.

Briefly, TMLEs are plug-in estimators that are asymptotically linear, often doubly robust, and semiparametrically efficient under standard regularity conditions \citep{van2011targeted}. Implementing a TMLE typically consists of two steps: (1) an initial estimate of the relevant portion of the likelihood (in this context, the conditional hazard of death at each time point) and (2) a targeting step that updates the initial estimate in order to optimize the bias–variance trade-off for the parameter of interest.

While TMLEs for the survival curve have been well studied in settings with classic right censoring \citep{brooks2013targeted, cai2020one}, the contribution here lies in extending the TMLE methodology to accommodate complex resampling design data and leveraging the distinct structure of the coarsening at random (CAR) likelihood for this problem. Specifically, two key features characterize this setting: the target parameter depends only on the conditional hazard of death and not on the conditional distribution of the time-dependent covariates $\bar{L}(t)$, and the structure of the CAR likelihood permits the use of both baseline and future covariate information ($W, \bar{L}(\tau)$) when estimating the hazard.

Together, these properties enable the use of point-treatment-style estimators that fully incorporate time-dependent covariate information (as if they were baseline covariates) while avoiding the need to model the conditional distributions of the time-dependent covariate $\bar{L}(t)$ terms, a difficult task which is often necessary in longitudinal data settings.

\subsection{Overview of currrent literature} \label{subsec:literature}

A growing body of work uses resampling (or double-sampling) designs to recover missing outcomes among LTFU patients and incorporate them into population-level survival estimates. Most commonly, researchers employ weighted Kaplan-Meier (KM) estimators, which assign: weight 1 to participants with known outcomes from clinic records; weight 0 to LTFU participants whose outcomes remain unknown after resampling; weight equal to the inverse of the resampling probability to LTFU participants who are subsequently chosen for resampling. These weighted KM estimators, used in settings such as Uganda, Kenya, Zambia, and across East Africa \citep{geng2015estimation, holmes2018estimated, yiannoutsos2008sampling}, are unbiased under known sampling probabilities and independence assumptions. However, they do not incorporate covariate information — even when survival up to time $t$ and time-varying covariates (e.g., CD4 count or viral load) are available. This leads to inefficient use of the data and can result in suboptimal inference, particularly when follow-up is limited or covariates are strongly predictive of survival.

\citet{yiannoutsos2008sampling} compared four mortality estimators and found that approaches accounting for outcomes among traced patients yielded estimates more than fivefold higher than unadjusted KM. Moreover, the authors found that weighted averages of survival functions (as in standard KM approaches) may fail to account for differential hazards across subgroups. This motivates alternative methods that operate directly on hazard functions.

\citet{frangakis2001addressing} propose such an approach, constructing survival curves by estimating the average hazard across dropout and non-dropout groups, weighted by group size. This avoids direct reliance on the Kaplan-Meier estimator, which they argue is inappropriate when administrative censoring is dependent on survival within groups. However, their estimator was later critiqued by \citet{robins2001discussion}, who showed that it is equivalent to an inefficient inverse probability weighted estimator in the full-data setting and proposed an IPCW-Nelson-Aalen estimator that achieves better asymptotic variance. \citet{li2008non} extended Nelson-Aalen estimation to this setting, constructing survival estimators that borrow strength across stages to reduce finite-sample bias and improve coverage. 

More recent semi-parametric work has focused on double-sampling for general coarsening structures, inspired by missing data structures that arise in the analysis of electronic health record data. \citet{levis2022double} derived efficient influence functions (EIFs) and one-step estimators for the causal average treatment effect (ATE) under minimal assumptions. Their estimators achieve efficiency and robustness, but do not explicitly handle censored survival outcomes or incorporate time-varying covariates. Their work does, however, provide a general roadmap for EIF-based estimation of the ATE under double sampling designs.

An alternative line of work, initiated by \citep{frangakis2015deductive}, proposes a deductive approach that avoids analytic derivation of the EIF in favor of numerical approximations using the Gateaux derivative. This method, later extended by \citep{qian2020deductive} to resampling settings with time-varying follow-up, enables inference without requiring closed-form expressions of the EIF. However, the deductive approach sacrifices analytical tractability and can obscure conditions for robustness, boundedness, and efficiency.

\section{Methods}\label{sec:methods}
Let $W$ represent covariates collected on study participants at baseline, $\tau$ be a participant's end of study (adminstrative censoring) time, and $\bar{L}(t)$ be time-dependent covariates collected on a participant through time $t$. Here, we use the overbar notation to denote the historical values of a given random variable collected through time $t$. $L(t)$ includes an indicator $V(t)$ of whether a patient makes a visit at $t$, an indicator $Id(t)$ of a participant having died and having had their death reported to the clinic at $t$, and $Lt(t)$ which is a vector of time-dependent biomarkers or covariates measured at $t$ if a patient makes a visit at that time $V(t)=1$ (thus we could more accurately write $Lt(t)=V(t)Lt(t)$ to denote that $L(t)$) will be NA if a visit is not made). Let $M$ represent the time at which a patient makes their last visit to the clinic. Let $T$ be time of death. Let $R$ be an indicator that a patient is chosen for resampling. Once resampled, we observe whether a patient has died before their end of study $\tau$ as well as their continuous time of death $T$ if applicable. Let $\Delta$ be an indicator that a patient has a known outcome at their end of study $\tau$. $\Delta$ is equal to 1 if a participant: (1) is known to be alive at their end of study $\tau$ because they make a visit at $\tau$ (ie $M=\tau$), (2) becomes LTFU, is chosen for resampling $R=1$, and is found to be alive at $\tau$, (3) becomes LTFU, is chosen for resampling $R=1$, and is found to be dead at or before $\tau$, or (4) is known to be dead at or before $\tau$ by having their death reported directly to the clinic $Id(t)=1$ during their study period (ie Id(t)=1 for any $t \leq \tau$). Let $\tilde{T}=\text{min}(T, \tau)$. If $\Delta=1$ outcome information is observed, namely $I(T \leq \tau)$ and $I(T \leq \tau)\tilde{T}$. If $\Delta=0$, we only know that $T>M$.

The full data $X=(W, \bar{L}(T), T)$ includes baseline covariates, time-dependent covariates tracked through $T$, and $T$ itself. This full data record stops when $T$ is observed. We have two censoring variables on top of this full data structure ($\tau$, $\Delta$). Let $\Delta^*=I(T \leq \tilde{T})$ meaning a patient dies before their end of study $\tau$. Given our two censoring variables ($\tau, \Delta$), our observed data can be written $O=\Phi(\tau, \Delta, X)=(W, \tau, \bar{L}(\tau), \Delta, \Delta \Delta^*, \Delta \Delta^* \tilde{T})$ and includes baseline covariates $W$, adminstrative censoring $\tau$, the covariate history tracked through a patient's end of study $\bar{L}(\tau)$, an indicator $\Delta$ of having known outcome, an indicator $\Delta^*$ of death happening before censoring, and time of death $T$ if observed before end of study $\Delta \Delta^* \tilde{T}$. Note that we only have covariate information on a participant up until the last visit time $M$ a participant makes, thus $\bar{L}(\tau)$ only contains meaningful information through $\bar{L}(M)$. If a participant misses a clinic visit at a given $t$, then no additional covariate information is measured at that time. 

Under coarsening at random (CAR), our censoring variables can only depend on the full data through the fully observed data $O$, ie $g(\tau, \Delta \mid X)=h(O)$ for some function $h$. In particular, the hazard of censoring by $\tau$ at time $t$ can depend on $(W, \bar{L}(t-))$, where $\bar{L}(t-)$ are covariates collected up until just before (but not at) $t$. CAR also implies that $P(\Delta=1 \mid \tau, X)=P(\Delta=1 \mid W, \tau, \bar{L}(\tau))$, the probability of having known outcome status $\Delta=1$ can depend on the entire covariate history through $\tau$. Conveniently, in this study design, the CAR assumption on $\Delta$ holds by design because researchers can only set resampling probabilities based on data they have collected by participant end of study $\tau$. The CAR assumption on $\tau$ also holds by design in the setting where total study duration $\tau$ only depends on a participant's observed study entry time or $\tau$ depends on other fully observed variables. We can write the CAR likelihood (determine the coarsening $C(o)$ of $X$ for each type of observations $o$) by considering what data is observed on what groups of participants in our study. The following groups emerge: (1) if $\Delta=0$, we observe covariate information but no information on the outcome $T$ (other than $T>M$), (2) if $\Delta=1$ and $\Delta^*=0$ we observe covariate information as well as information that a participant lived through their end of study $T > \tau$, and (3) $\Delta=1$ and $\Delta^*=1$ we observe covariate information and a participant's time of death $T$ which occurred before their end of study $\tau$.

These three groups correspond to the three distinct portion of our CAR likelihood below. Let $\lambda(s \mid W, \bar{L}(\tau))$ be the conditional hazard of $T$ at time $s$. 
\begin{align*}
    p_X(W, \tau, \bar{L}(\tau), \delta, \delta \delta^*, \delta^* \tilde{t}) 
    &= I(\delta=0)q_W(W)P(\bar{L}(\tau) \mid W) \\
    &\quad + I(\delta=1, \delta^*=0)P(\bar{L}(\tau) \mid W) \\
    &\times \prod_{s \leq \tilde{t}} (1-\lambda(s \mid W, \bar{L}(\tau))) \\
    &\quad + I(\delta=1, \delta^*=1)P(\bar{L}(\tau) \mid W) \lambda(\tilde{t} \mid W, \bar{L}(\tau)) \\
    &\times \prod_{s < \tilde{t}} (1-\lambda(s \mid W, \bar{L}(\tau))).
\end{align*}

The form of the conditional hazard of death $\lambda(s \mid W, \bar{L}(\tau))$ in this likelihood allows the hazard of dying at $s$ to depend on the entire $(W, \bar{L}(\tau))$ process. This implies that we can (and should) leverage \textit{future} covariate information when fitting the conditional hazard at time $s$. For example, we may use that the conditional hazard of death is deterministically 0 at every $t \leq M$. With long study periods, this gives the analyst access to vast amounts of powerful, highly predictive covariate information if leveraged by the chosen estimator. This parameterization of the likelihood is correct when $\tau$ is fixed and all partcipants have equal follow-up time. However, when $\tau$ is random, this parameterization presents a problem described further in varied follow-up $\tau$ section~\ref{subsec:varied_tau}. 

We propose two estimators for survival in resampling studies: (1) a semiparametric efficient TMLE for fixed follow-up designs, and (2) an IPCW-TMLE that accommodates varied follow-up times. Both estimators rely on analytic influence curve derivation and flexibly incorporate time-varying covariates through Super Learner, yielding estimators that are doubly robust fully efficient in some settings.

The rest of this section is organized as follows. First, in section ~\ref{subsec:fix_tau}, we introduce an efficient TMLE for survival in fixed follow-up resampling designs, where all participants have equal amounts of study follow-up $\tau$. Then, in section  ~\ref{subsec:varied_tau}, we introduce a TMLE for the varied follow-up case, where participants may have different amounts of study follow-up $\tau$. We begin by presenting a TMLE for the varied follow-up case which relies on stratification on participants with sufficient follow-up $\tau$ to evaluate survival (section ~\ref{subsubsec:strat_tmle}). We then present an IPCW-TMLE for the varied follow-up case which can be made highly efficient by estimating and/or targeting the follow-up censoring mechanism $\tau$ (sections ~\ref{subsubsec:ipcw_tmle}). 

\subsection{Fixed Follow-up Case ($\tau$) TMLE}\label{subsec:fix_tau}
We begin by constructing an estimator  for the setting in which each participant has a fixed amount of follow-up time (say $\tau=10$ months for everyone). Our target parameter is survival at some time of interest $t_0$ (alternatively, the entire survival curve through $t_0$). 

This survival probability can be expressed as the expectation, over $(W, \bar{L}(\tau))$, of the conditional probability of surviving beyond $t_0$, given $W$, $\bar{L}(\tau)$, and $\Delta=1$. By invoking the coarsening at random (CAR) assumption, $\Delta \perp T \mid W, \bar{L}(\tau), \tau$, we may condition on $\Delta=1$ directly. Note that we only observe right censored (by $\tau$) event times $\tilde{T}$ for individuals with $\Delta=1$. 
$\tau$ cannot be treated as a right censoring time for particiapnts with $\Delta=0$ since we do not know if a patient dies between $M$ and $\tau$. Applying the product integral representation of the survival function, we write:
\begin{eqnarray*}
    \mathbb{E}_{W,\bar{L}(\tau)}P(T > t_0 \mid W, \bar{L}(\tau), \Delta=1) = \mathbb{E}_{W,\bar{L}(\tau)} \prod_{s \leq t_0}(1-\lambda(s \mid W, \bar{L}(\tau), \Delta=1))
\end{eqnarray*}
The form of this target parameter motivates the construction of a Targeted Maximum Likelihood Estimator (TMLE) which estimates and targets the conditional hazard of death using observations with $\Delta=1$ and subsequently imputes hazard values for participants with $\Delta=0$. By using the product integral representation, the resulting plug-in estimator naturally respects the monotonicity of the survival function.

\subsubsection{Estimation Procedure}\label{subsubsec:fix_tau_est_proc}
Targeted maximum likelihood estimators involve first obtaining an initial estimate of the relevant portion of the likelihood, in this case, the conditional hazard of death at each time point $t \leq t_0$. We could opt to fit the conditional hazard function at each time point separately, but we suggest using a pooled estimator which fits the conditional hazard of death across all time points $t \leq t_0$ simultaneously. This can be done by creating a repeated measured data structure with rows $(M+1):min(T, \tau)$ for participants with $\Delta=1$ and fitting a regression of the form:
$$dN(t) \sim t + W + \bar{L}(\tau)$$
where $dN(t) = \mathbb{I}(T = t)$ is the increment of the counting process. Here, we use $\sim$ to denote regression.

When fitting this regression, we exclude repeated measure observations at time points $t \leq M_i$ for each person $i$ because individuals are not considered at risk of death during periods in which they are still attending clinic visits (the hazard is deterministically zero in this region). A key strength of our approach lies in its use of the entire time-dependent covariate process $\bar{L}(\tau)$ in modeling the hazard (in particular, making use of deterministic information by plugging in known values for the hazards when appropriate, as we do with hazards at $t \leq M$).  We recommend fitting this hazard regression using Super Learner, an ensemble machine learning method that flexibly combines multiple candidate learners to reduce reliance on restrictive modeling choices \citep{van2007super}. Once fit, the model is used to predict the conditional hazard of death for all individuals (including those with $\Delta=0$) at each time point from $M+1$ through $t_0$, where the participant is considered at risk of death.

The second step of the TMLE procedure, ``the targeting step," updates the initial hazard estimates to achieve an optimal bias-variance tradeoff for the target parameter and ensure valid inference \citep{van2011targeted}. This step involves a logistic regression of the observed outcome $dN(t)$ on a special covariate known as the ``clever covariate", denoted $H(t)$, with the initial estimate $\hat{\lambda}(t \mid W, \bar{L}(\tau))$ included as an offset: 
\begin{align}
dN(t) \sim \text{offset(plogis}(\hat{\lambda}(t \mid W, \bar{L}(\tau)))) + H(t)
\end{align}
The clever covariate $H(t)$ is derived from the efficient influence function (EIF) of the target parameter, which takes the following form for the fixed follow-up $\tau$ case:
\begin{align*}
    D^*_{\tau}(P) &= \sum_{t=1}^{\tau} I(T \geq t) \left( I(t \leq t_0)  \frac{I(\Delta=1)}{P(\Delta=1 \mid W,\bar{L}(\tau))}  \frac{S(t_0 \mid W, \bar{L}(\tau))}{S (t \mid W, \bar{L}(\tau))}  \right)\\
    & \times (dN(t) - \lambda(t \mid W, \bar{L}(\tau)))  + S(t_0 \mid W, \bar{L}(\tau)) - \Psi(P)
\end{align*} 
The derivation of this efficient influence curve follows exactly from the derivation of the EIF for right-censored survival data established previously \citep{van2011targeted} with one modification: we replace baseline covariates $W$ with $(W,\bar{L}(\tau))$. Recall, our ability to treat time-dependent covariates as baseline covariates in this setting is justified by the form of the coarsening at random (CAR) likelihood for this resampling design application. 

The clever covariate, $H(t)$, corresponds to the portion of the EIF that involves the resampling (or more generally, exposure) mechanism:
\begin{align*}
    H(t) &= I(T \geq t) \left( I(t \leq t_0)  \frac{I(\Delta=1)}{P(\Delta=1 \mid W,\bar{L}(\tau))}  \frac{S(t_0 \mid W, \bar{L}(\tau))}{S (t \mid W, \bar{L}(\tau))}  \right)
\end{align*} 
At each time point $t$, $H(t)$ is computed for each individual using the initial hazard estimates for the survival terms and known probabilities $P(\Delta = 1 \mid W, \bar{L}(\tau))$.

Following the targeting step, we obtain updated estimates of the conditional hazard:
$$\hat{\lambda}^*(t \mid W, \bar{L}(\tau)) = plogis(qlogis(\hat{\lambda}(t \mid W, \bar{L}(\tau)))) + \hat{\epsilon}*H(t)$$
where $\hat{\epsilon}$ is the estimated coefficient in front of $H(t)$ as estimated during the targeting regression. The superscript $^*$ denotes the targeted (updated) estimate of the hazard. The goal of this step is to ensure that the efficient influence function is mean zero under the estimated distribution, i.e., $\mathbb{E}[D^*_{\tau}(\hat{P})] = 0$, thereby eliminating first-order plug-in bias \citep{van2011targeted}.

Once the conditional hazard estimates for each time point $t \leq t_0$ have been estimated and targeted for each participant, they are plugged into the product integral representation of the survival curve. For time points where a participant's hazard is deterministically known, known values are substituted directly, bypassing the need for estimation or targeting. In many cases, the majority of a participant’s conditional hazards are entirely determined by their covariate data, further enhancing the efficiency and stability of this estimator.

\subsubsection{Targeting Options}\label{subsubsec:targeting_options}

We outline two valid approaches to implement the targeting step for the conditional hazard in this context: pooled targeting and recursive targeting. Both methods ultimately solve the same efficient score equation but differ in their implementation and, potentially, their practical performance.

\noindent
\textbf{Pooled Targeting}\\
In the pooled approach, the targeting step is performed by applying the logistic regression model in Equation (1) to the full repeated-measures data structure, updating the conditional hazard of death at all time points $t \leq t_0$ simultaneously. This method allows us to leverage all available data at once, essentially ``borrowing" information across time.

The clever covariate $H(t)$ depends on the conditional hazards themselves, through the ratio of survival probabilities. As a result, this targeting step must be applied iteratively. After each targeted update of the hazard estimates, the clever covariate and the efficient influence curve (EIF) must be recomputed using the updated estimates $\hat{\lambda}^*(t \mid W, \bar{L}(\tau))$. Iteration continues until convergence, defined as the point at which the empirical mean of the EIF is sufficiently close to zero. We recommend the following convergence criteria: $\frac{1}{n}\sum_{i=1}^n \hat{D}^*_{\tau}(P) \leq \frac{\sqrt{var(\hat{D}^*_{\tau}(P))}}{\sqrt{n}log(n)}$.This ensures that the efficient score equation $\mathbb{E}[D^*_{\tau}(P)] = 0$ is approximately solved, thereby removing first-order bias. \\
\noindent
\textbf{Recursive Targeting}\\
An alternative strategy is recursive targeting, which avoids iteration by targeting the conditional hazard each time point sequentially rather than simultaneously. The procedure begins by targeting the conditional hazard at the final time point $t = t_0$ and proceeds backward in time until the conditional hazard at $t = 1$ has been targeted.

At $t = t_0$, the survival ratio in $H(t)$ simplifies (since $\frac{S(t_0 \mid W, \bar{L}(\tau))}{S (t \mid W, \bar{L}(\tau))} = 1$ when $t = t_0$), so the score equation for that time point can be solved in a single step. For earlier time points $t < t_0$, the survival ratio will depend only hazards at future time points ($s > t$), which have already been targeted. Thus, the updated clever covariate at time $t$ depends only on previously updated hazard estimates, avoiding the need for iteration. This can be seen by expressing the survival terms in $H(t)$ as products of conditional hazards and observing that all terms cancel in the ratio except those with $s > t$. \\
\noindent
\textbf{Targeting Recommendation: Pooled Targeting}\\
Although both targeting approaches solve the same efficient influence equation, we recommend the pooled targeting approach for applications involving resampling designs. In traditional survival settings, data become sparse at later time points as fewer individuals remain uncensored. In our setting, an additional challenge arises: data are also sparse early in follow-up because individuals with $\Delta=1$ only contribute to hazard estimation when $t > M$, leaving limited support at early times when most participants are known to be alive. Taken together, this leads to instability in recursive targeting. Pooled targeting mitigates this issue by borrowing strength across time, smoothing over regions with limited data and producing more stable updates to the hazard estimates.

\subsection{Varied Follow-up ($\tau$) Case TMLE} \label{subsec:varied_tau}
We now turn to the more general setting in which participants have heterogeneous follow-up durations. Such variation may arise, for example, due to staggered entry into the study. Let $\tau$ denote each individual’s amount follow-up time, which we assume is drawn from the interval ${1, \ldots, t_{\max}}$, where $t_{\max}$ is the maximum possible follow-up time.

Recall that the CAR (coarsening at random) likelihood involves the conditional hazard term $\lambda(t \mid W, \bar{L}(\tau))$. In the fixed follow-up setting, where all participants have the same $\tau$, this quantity corresponds to a common conditional distribution across individuals. However, in the varied $\tau$ setting, this term depends explicitly on each individual's unique follow-up history $\bar{L}(\tau)$, resulting in a different conditional distribution for each value of $\tau$.

For example, the conditional hazards $\lambda(t \mid W, \bar{L}(4))$ and $\lambda(t \mid W, \bar{L}(5))$ are not independent objects; they are linked through the underlying conditional event-time distribution. In particular, by the law of total probability:
\begin{align*}
    P_{T}(t \mid W, \bar{L}(4)) = \sum_{l(5)} P_T(t \mid W, \bar{L}(5)=l(5)) P(\bar{L}(5)=l(5) \mid \bar{L}(4), W).
\end{align*}
Hazards inherit this structural relationship indirectly, since each hazard is defined from its corresponding conditional density and survival function. This identity highlights a key structural property of the likelihood: the conditional distribution at a shorter history length (e.g., $\bar{L}(4)$) is a marginalization over the longer one (e.g., $\bar{L}(5)$). Ideally, estimation should honor this structure.

However, if we separately fit $P_T(t \mid W, \bar{L}(4))$, $P_T(t \mid W, \bar{L}(5))$, and $P(\bar{L}(\tau) \mid W)$, as this factorization might suggest, there is no guarantee that the above identity will hold—especially in finite samples. This leads to a misspecified or incoherent parametrization, in which likelihood components are estimated independently despite being functionally related. As a result, the estimator may discard valid information embedded in the joint structure of the data and suffer from (1) a loss in efficiency or (2) incompatibility with the actual data density. A maximum likelihood estimator which ignored this structure would be inefficient or result in a non-probability measure. While we do not resolve this issue here, we highlight it as an important caveat and a direction for future work.

Before introducing the stratified and IPCW-TMLE estimators we propose for the varied follow-up $\tau$ setting, we briefly explain why we adopt this approach rather than constructing a fully efficient estimator, which would be preferable in principle. The fundamental challenge is that the bivariate censoring structure induced by $(\tau, \Delta)$---where $\tau$ censors the covariate trajectory $\bar{L}(t)$ and $\Delta$ censors the event process---does not seem to admit a closed-form efficient influence function for our target parameter. The difficulty of estimation with bivariate censored data has been extensively presented elsewhere \citep{laan2003unified}. Without an explicit EIF, we cannot implement the standard TMLE targeting procedure. An alternative approach would be to construct a maximum likelihood estimator by fitting the entire CAR likelihood under both censoring mechanisms. However, this path also encounters substantial obstacles. Even in the special case where $L(t)$ reduces to a single time-dependent indicator, the nonparametric likelihood for this joint censoring structure does not admit a closed-form NPMLE. Obtaining the MLE would require an Expectation-Maximization (EM)-type algorithm over the longitudinal data structure with censoring. While using, for instance, the Highly Adaptive Lasso (HAL) within such an EM framework is theoretically possible and would respect the CAR likelihood structure, this approach is operationally demanding and requires estimating all full-data conditional distributions of the $L(t)$ nodes given the past. Moreover, inference would still require the nonparametric bootstrap without access to the closed-form EIF. For these reasons, we instead adopt an IPCW-TMLE approach. We apply our proposed efficient TMLE developed for the fixed-$\tau$ case (which handles censoring by $\Delta$) but augment it with additional inverse-probability weights to handle the second censoring mechanism induced by $\tau$. Although this estimator is not fully efficient in the full bivariate censoring setting, it provides a coherent and practically implementable procedure. We do, however, provide augmentations to this IPCW-TMLE, such as flexibly estimating and targeting the IPCW weights themselves, that yield a highly efficient estimator in practice.

\subsubsection{Stratified TMLE} \label{subsubsec:strat_tmle}

We propose adapting our fixed $\tau$ TMLE by (1) using an estimator for the initial fit of the conditional hazards which uses all data and (2) performing the targeting step and evaluation of survival among the subset of data for which $\tau \geq t_0$. This stratified TMLE yields an unbiased estimate of the marginal survival probability at time $t_0$ under the assumption that follow-up time $\tau$ is completely random and therefore independent of the failure time $T$, ($P[T > t_0 \mid \tau_i]=P[T > t_0]$). Under this assumption, the target parameter can be rewritten as:
$$S(t_0) = \mathbb{E}_{\tau}[P[T>t_0 \mid W, \bar{L}(\tau)] \mid \tau \geq t_0]$$
By using the full sample for the initial hazard estimation, including individuals with $\tau < t_0$, we aim to maximize efficiency by leveraging the entire observed dataset for this first step. Restricting only the targeting and evaluation steps preserves the integrity of the parameter definition while maintaining as much information as possible in the nuisance estimation.

Importantly, when all individuals satisfy $\tau \geq t_0$, this stratified TMLE coincides with the fully efficient estimator developed for the fixed follow-up setting. However, when a substantial proportion of participants have $\tau < t_0$, the procedure becomes inefficient due to reduced sample size in the targeting and evaluation steps. This trade-off highlights a key limitation of stratification: while it maintains validity under random follow-up, efficiency may deteriorate when early end of study is common.

The influence curve of this estimator is equal to that of the EIF for the fixed $\tau$ case, $D^*_{\tau}(P)$, with the addition of weights. 
\begin{align*}
    D^*_{G}(P) &= \frac{I(\tau \geq t_0)}{\bar{G}(t_0 \mid W, \bar{L}(t_0)))}D^*_{\tau}(P)
\end{align*}
where $\bar{G}(t_0 \mid W, \bar{L}(t_0))$ is the conditional survival probability for being uncensored through $t_0$, given baseline and the history of time-dependent covariates leading up to $t_0$. Specifically,
$$\bar{G}(t_0 \mid W, \bar{L}(t_0))) = \prod_{s \leq t_0}(1-\lambda_{\tau}(s \mid W, \bar{L}(s-)))$$
with $\lambda_{\tau}(s \mid W, \bar{L}(s-))$ denoting the conditional hazard of being censored by time $s$. In this application, these weights are known by design as $\tau$ is completely determined by a participants' observed entry time into the study. These known censoring probabilities can be directly plugged in to the above representation to adjust the inference accordingly.

\subsubsection{IPCW-TMLE}\label{subsubsec:ipcw_tmle}
To recover efficiency lost due to stratification—or to accommodate informative follow-up times $\tau$—we propose an Inverse Probability of Censoring Weighted TMLE (IPCW-TMLE) with estimated weights. This approach introduces inverse probability weights to either gain efficiency by estimating weights when they are known or to account for informative censoring when $\tau$ is not fully, but conditionally, independent of the failure time $T$.

In general, an IPCW-TMLE is constructed by applying a standard TMLE to some full data structure $X$, with the addition of inverse weights to address censoring \citep{rose2011targeted}. Let us re-define the ``full data" as:
$$X=(W, \bar{L}(t_{0}), \Delta, \Delta \Delta^*, \Delta \Delta^*T)$$. We treat $\tau$ as a censoring variable on top of $X$, yielding the observed data:
$$O=(W, \tau, \bar{L}(\tau), \Delta, \Delta \Delta^*, \Delta \Delta^* \tilde{T})$$
The IPCW-TMLE proceeds by applying the fixed-$\tau$ TMLE to the full data structure $X$, with the inclusion of censoring weights
$$\frac{I(\tau \geq t_0)}{\bar{G}(t_0 \mid W, \bar{L}(t_0)))}$$
as defined previously. This IPCW-TMLE differs from the Stratified TMLE presented above in that the weights are not only used to get correct inference but included during the point estimation procedure. This formulation can be extended to allow $\tau$ to depend on observed covariates, thereby adjusting for informative censoring. In either case, estimation of the conditional hazard of censoring can be conducted flexibly using machine learning methods. These estimated weights are then used during the targeting step for the conditional hazard of death (ie the targeting step is now a \textit{weighted} logistic regression). 

The influence function for the IPCW-TMLE estimator with known weights takes the same form as for the Stratified TMLE:
\begin{align*}
    D^*_{G}(P) &= \frac{I(\tau \geq t_0)}{\bar{G}(t_0 \mid W, \bar{L}(t_0)))}D^*_{\tau}(P)
\end{align*}
where $D^*_{\tau}(P)$ is the EIF corresponding to the fixed follow-up case.

However, when the censoring mechanism is estimated even though it is known by design, the influence curve is no longer simply $D^*_{G}(P)$ but $D^*_{G}(P)$ minus its projection onto the tangent space of the censoring mechanism $T_{CAR}=T_G(P)$. Flexible or targeted estimation of the follow-up censoring mechanism $\tau$ when the true weights are known can make this estimator highly efficient despite the required stratification on $\tau \geq t_0$. Lemma~\ref{lem:IC-IPCW} provides the influence curve for the IPCW-TMLE with estimated weights. The full derivation of this IF can be found in Section~\ref{subsubsec:projection} of the Supplementary Material. 
\begin{lemma}\label{lem:IC-IPCW}
The influence function for the IPCW-TMLE with estimated weights (censoring hazard) is given by:
\begin{align*}
    D^{**}_G(t) &= D^*_G(t) - \prod(D^*_G \mid T_G(P)) \\
    &= \left[ \mathbb{E}\big(D^*_G(t) \mid \tau = t, \operatorname{Pa}(A(t))\big) 
    - \mathbb{E}\big(D^*_G(t) \mid \tau > t, \operatorname{Pa}(A(t))\big) \right] \\
    & \quad \times \big( dA(t) - \lambda_{\tau}(t \mid W, \bar{L}(t)) \big).
\end{align*}
\end{lemma}
Where $T_G(P)$ is the tangent space of the censoring distribution and $Pa(A(t))=(W, \bar{L}(t), A(t-1))$ are the parents of $A(t)$, the covariate history relative to the censoring time $t$.

\subsubsection{Targeting of G} \label{subsubsec:targeting_G}
Efficiency in this IPCW-TMLE can be improved through either flexibly estimating and/or targeting the inverse censoring weights. We now demonstrate the targeting procedure for these inverse weights. Let $f(t)$ denote the clever covariate used in the targeting step for the censoring mechanism ($f(t)$ comes from the projection of $D^*_G$ onto the tangent space of the censoring mechanism):
\begin{align*}
    f(t) &=  \Big[ E(D^*_G \mid dA(t) = 1, Pa(A(t))) - E(D^*_G \mid dA(t) = 0, Pa(A(t))) \Big]
\end{align*}
The targeting step for $G$ involves a logistic regression updating step with $f(t)$ as the covariate and the initial estimate of the conditional hazard of censoring included as an offset:
$$dA(t) \sim \text{offset(plogis}(\hat{\lambda}_{\tau}(t \mid W, \bar{L}(t-)))) + f(t)$$
Following this update, the conditional hazard of censoring $\hat{\lambda}_{\tau}(t \mid W, \bar{L}(t-)))$ is recalculated in a similar fashion as described for the outcome regression targeting. This targeting step aims to make the empirical mean of the following term approximately zero:
\begin{align*}
    & \sum_t f(t) \Big( dA(t) - \lambda_{\tau}(t \mid W, \bar{L}(t)) \Big).
\end{align*}
which reflects the efficient score equation for $G$. As in the outcome targeting step, this procedure is iterative, since $f(t)$ depends not only on the current estimate of the censoring hazard but also on the conditional hazard of the outcome through the targeting step's reliance on $D^{*}_G$. Convergence of the targeting step is achieved when the empirical mean of the influence function $D^{**}_G$ is sufficiently close to zero, using a stopping criterion analogous to that used in the outcome TMLE step.

Regardless of whether we estimate $G$ flexibly or apply a targeting step, the goal is the same: to reduce the variance of the resulting estimator. Evidence of improved efficiency can be shown if $var(D^{**}_G) < var(D^*_G)$.

In order to implement this targeting step for the censoring mechanism $G$, one needs to estimate the projection term $f(t)$ using a regression method of choice. The entire TMLE procedure then requires a joint targeting procedure which iterates back and forth between targeting the outcome regression (the conditional hazard of death) and the censoring mechanism (the conditional hazard of censoring). Details on estimating the projection term are provided in 
Section~\ref{subsubsec:est_proj} of the Supplementary Material. The full iterative targeting algorithm is described in
Section~\ref{subsubsec:joint_targeting} of the Supplementary Material.

\section{Simulations \& Results}\label{sec:sims}

In this section, we present simulation studies to assess the performance of our proposed TMLEs relative to estimators more commonly used in the resampling design literature. In section~\ref{subsec:data_gen}, we describe the data generating process used for both the fixed $\tau$ and varied $\tau$ simulations. Then, in section~\ref{subsec:estimator_comp} we describe the comparator estimators from the literature. Finally, in section~\ref{subsec:sim_results}, we will provide simulation results comparing bias, variance, mean squared error (MSE), and 95\% confidence interval coverage of these estimators.

\subsection{Data generating processes}\label{subsec:data_gen}
We simulated longitudinal data for $n = 3000$ individuals over $t = 10$ time points. This data includes: baseline covariates, time-varying visit indicators, time-varying CD4 biomarker values (collected only if a visit is made at that $t$), time-varying death indicators (potentially unobserved), time-varying natural death reporting indicators (if a patient dies and has death reported to the clinic on their behalf), and a resampling indicator which triggers the uncovering of (otherwise unobserved) death information if a participant is resampled. Baseline covariates included three binary variables ($W_1, W_2, W_3$) representing risk and protective factors. At each time point, individuals faced a probability of death dependent on their baseline covariates, prior visit behavior, and prior CD4 count values. If death occurred (potentially unobserved) time of death $T$ was generated. If a death occurred, it was naturally reported to the clinic $\text{Id}(t)$ with probability 0.2 among newly deceased individuals, remaining constant thereafter. Visit $V(t)$ probabilites were drawn based on covariates, past visit history, and lagged CD4 counts, with biomarker values $L(t)$ representing CD4 levels evolving autoregressively with Gaussian noise and penalized means based on covariates and visit behavior. Observed CD4 counts were carried forward from previously observed values if a visit was not made at a given time. Each individual was assigned a censoring time $\tau \in \{5,7,9,10\}$ with different probability for each $\tau$, and all data past $\tau$ were censored. The outcome was an indicator of death by time $t_0$, derived either from observed clinic deaths, observed clinical visits (indicating survival), or resampling. Individuals with unobserved outcomes were chosen for resampling (and subsequent outcome measurement) with probability 0.2. In simulation studies for the ``fixed follow-up ($\tau$) case," $\tau$ was set to 10 for all participants (this was the only deviation from the above procedure required). Further information and data generating simulation code can be found in the Supplementary material~\ref{subsubsec:sim_setup}.

\subsection{Estimators}\label{subsec:estimator_comp}

We now detail all comparator estimators used in both the fixed-$\tau$ and variable-$\tau$ simulation studies, along with the procedures used to compute inference for each. Let $\Pi_{\Delta}=P(\Delta=1 \mid W, \bar{L}(\tau))$ and $\Pi_{\tau}=\lambda_{s}(s \mid W, \bar{L}(s-))$.

\subsubsection{Fixed $\tau$ case}

The following estimators were compared for the fixed $\tau$ case
\begin{enumerate}
    \item TMLE (pooled)
    \item Untargeted Plug-in
    \item Inverse probability weighting (IPW) with known weights $\Pi_{\Delta}$
    \item Inverse probability weighting (IPW) with estimated weights $\Pi_{\Delta}$
    \item Weighted Kaplan-Meier (wKM) with known weights $\Pi_{\Delta}$
    \item Weighted Kaplan-Meier (wKM) with estimated weights $\Pi_{\Delta}$
    \item Naive Kaplan-Meier (KM)
\end{enumerate}

The first estimator (1) TMLE (pooled) is the TMLE presented in section~\ref{subsec:fix_tau} with the pooled targeting step. Recall with fixed $\tau$, we have sufficient follow-up on all participants to evaluate survival at all ten time points, ie $\tau \geq t_0$. For this simulation, known resampling probabilities $\Pi_{\Delta}$ were used in the denominator of the clever covariate. In other worlds, if a participant had a known outcome by their end of study $\tau$, $\Pi_{\Delta}=1$ deterministically, and if a participant had unknown outcome by $\tau$, then $\Pi_{\Delta}=0.2$, ie the known resampling probability. Inference for this pooled TMLE is computed via this estimators influence curve, the form of which is presented in section~\ref{subsubsec:fix_tau_est_proc}. 

The untargeted plug-in estimator (2) estimates the conditional hazard probabilities and plugs these estimates directly into the product integral representation of the survival curve without performing the targeting step. This estimator is equivalent to running our TMLE estimator (1) without the targeting step. Since Super Learner was used to generated initial predictions of the conditional hazard of death, there is no formal theory which supports valid inference for this estimator. 95\% confidence interval coverage is thus not reported, though the variance of the point estimate across simulation studies is reported.

The IPW estimator (3) estimates the survival probability by time $t_0$ by weighting the observed survival indicator from each participant with $\Delta=1$ by the inverse of their known resampling probability. This estimator ignores all data on participants with $\Delta=0$, a known source of inefficiency, but re-weights the data in order to account for this source of informative missingness. Known weights $\Pi_{\Delta}$ are used in the same manner described for our TMLE estimator (1). Inference for this estimator is reported via its influence curve. 

The IPW estimator (4) is the same as (3), but it estimates the inverse weights $\Pi_{\Delta}$ even though they are known by design. This estimator may gain efficiency over estimator (3) by incorporating data from participants with $\Delta=0$ via the estimation of the inverse weights. A logistic regression model for $\Pi_{\Delta}$ including baseline covariates, $M$ (time of last clinic visit attended), and the last measured value of $L(t)$ (our time-dependent biomarker variable) is fit using data from all participants who have non-deterministic values for $\Delta$ (essentially fitting the resampling probability among those eligible for resampling). Influence curve based inference is reported, though it may be conservative in cases where weight estimation markedly improves efficiency.

Estimator (5) is the weighted Kaplan-Meier estimator which is commonly used in resampling design literature. The weighted KM estimator gives a weight of 1 to observations with known outcome via clinical data, ie known alive via making visits or known dead via natural death reporting to the clinic. It gives a weight of 0 to observations with unknown outcome even after resampling. It gives a weight inverse to the probability of resampling to any observations that had and an unknown outcome and were subsequently chosen for resampling, thus uncovering their outcome value. Known resampling weights are used. We implement this estimator using the \texttt{survival} package in R \citep{survival-package}; however, inference for this estimator must be reported via bootstrapping. Note when applying KM in this fixed-tau setting, there is no censoring by $\tau$. 

Estimator (6) corresponds to the weighted Kaplan–Meier estimator in (5), but replaces the known weights with estimated weights obtained as in estimator (4). Although this approach does not appear to be widely adopted in the resampling design literature, similar to the IPW estimator, it has the potential to substantially reduce variance relative to using known weights.

The Naive KM estimator (7) demonstrates what survival estimates would look like without resampling. This estimator treats the last visit a participant makes to the clinic as their right censoring time and runs KM without including deaths discovered via resampling. Inference for the naive Kaplan-Meier estimator is computed with Greenwood's formula within the \texttt{survival} package in R \citep{survival-package}.

For estimators (1) and (2), we use Super Learner to obtain initial estimates of the conditional hazards at each time point (\texttt{SuperLearner} package in R \citep{Polley2023}). Because the effective sample size decreases after excluding observation–time pairs with deterministic hazards, we restrict the Super Learner library to linear and lasso regression models with varying subsets of past covariates. In our simulations, this choice provided a favorable balance between flexibility and stability. Incorporating more flexible learners or larger covariate sets tended to overfit in this setting, thereby diminishing the benefit of the targeting step and degrading performance. In applied analyses with larger sample sizes (and less deterministic structure), a more flexible library would be preferable to avoid reliance on restrictive parametric assumptions.

\subsubsection{Varied $\tau$ case}

The following estimators were compared in the simulations for the varied $\tau$ case. 
\begin{enumerate}
    \item IPCW-TMLE (pooled)
    \item Untargeted Plug-in of IPCW-TMLE
    \item IPCW-TMLE (pooled), estimated weights $\Pi_{\tau}$
    \item IPCW-TMLE (pooled), estimated weights $\Pi_{\tau}$ and $\Pi_{\Delta}$
    \item Weighted Kaplan-Meier, known weights $\Pi_{\Delta}$
    \item Weighted Kaplan-Meier, estimated weights $\Pi_{\Delta}$
    \item Naive Kaplan-Meier (KM)
\end{enumerate}

The IPCW-TMLE estimator (1) we apply is the one defined in section~\ref{subsubsec:ipcw_tmle}. We use known $\Pi_{\Delta}$ and $\Pi_{\tau}$ for this estimator. Again, this is a realistic choice given that, in our applied data application, the values of $\tau$ are known at baseline given a participant's entry time into the study an resampling probabilities are set by the study designers. 

The untargeted plug-in (2) again is the IPCW-TMLE estimator (1) without the targeting step. 

The IPCW-TMLE (3) is similar to (1) but the hazard of censoring by $\tau$, $\lambda_{\tau}(t \mid W, \bar{L}(t-))$, is estimated at each time point separately (ie stratified) using Super Learner. The covariates passed into Super Learner include the baseline covariates, a visit summary variable which represents the total number of visits made by a participant before $t$, and the most recent non-zero value of the time-dependent covariate $L(t)$ measured before $t$. The Super Learner library includes logistic regression, lasso regression, Bayes logistic regression, and multivariate adaptive regression splines for a more flexible fit.

The IPCW-TMLE (4) is similar to (3) in that the hazard of censoring by $\tau$ is estimated in the same manner. Additionally, $\Pi_{\Delta}$ is estimated among participants with non-determinsitic values of $\Delta$ given their covariate history. $\Pi_{\Delta}$ shows up in the clever covariate and, more generally, the influence curve itself. This regression of $\Delta$ on covariates is run (among those with unknown outcome at $\tau$) using Super Learner with baseline covariates, the time of last visit made to the clinic $M$, a participant's end of study  $\tau$, and the latest non-zero value of the time-dependent covariate $L$. The Super Learner library includes logistic regression, lasso regression, and multivariate adaptive regression splines to allow for a more flexible fit. 

Estimator (5) is the weighted Kaplan-Meier estimator with known weights which appears to be the common choice of estimator in applied resampling design literature. This estimator was described in the previous section. 

Estimator (6) is the weighted Kaplan-Meier estimator with estimated weights as described previously.

Estimator (7) is the naive Kaplan-Meier estimator as described prevously.

\subsection{Simulation results}\label{subsec:sim_results}

\subsubsection{Fixed Follow-up ($\tau$) Case Results}

\begin{figure}[H]
  \centering
  \includegraphics[width=\textwidth]{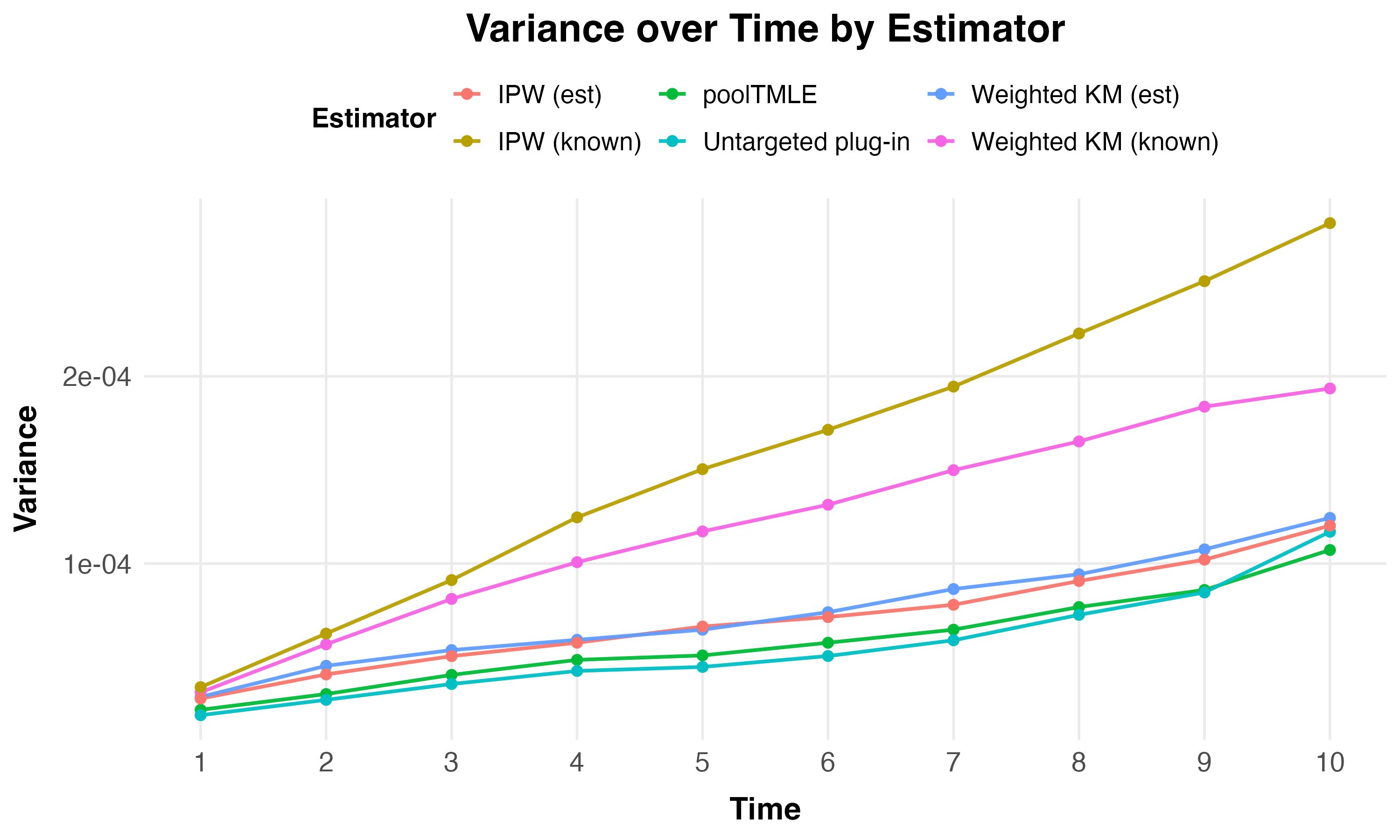}  
\caption{Variance of unbiased estimators over time in the fixed $\tau$ setting. The naive Kaplan-Meier estimator was omitted due to known bias in this setting. Here, (est) refers to estimated $\Pi_{\Delta}$ and (known) refers to the use of known values of $\Pi_{\Delta}$.}
  \label{fig:fixtau_graph}
\end{figure}

\begin{sidewaysfigure}
  \centering

  \includegraphics[width=\textheight]{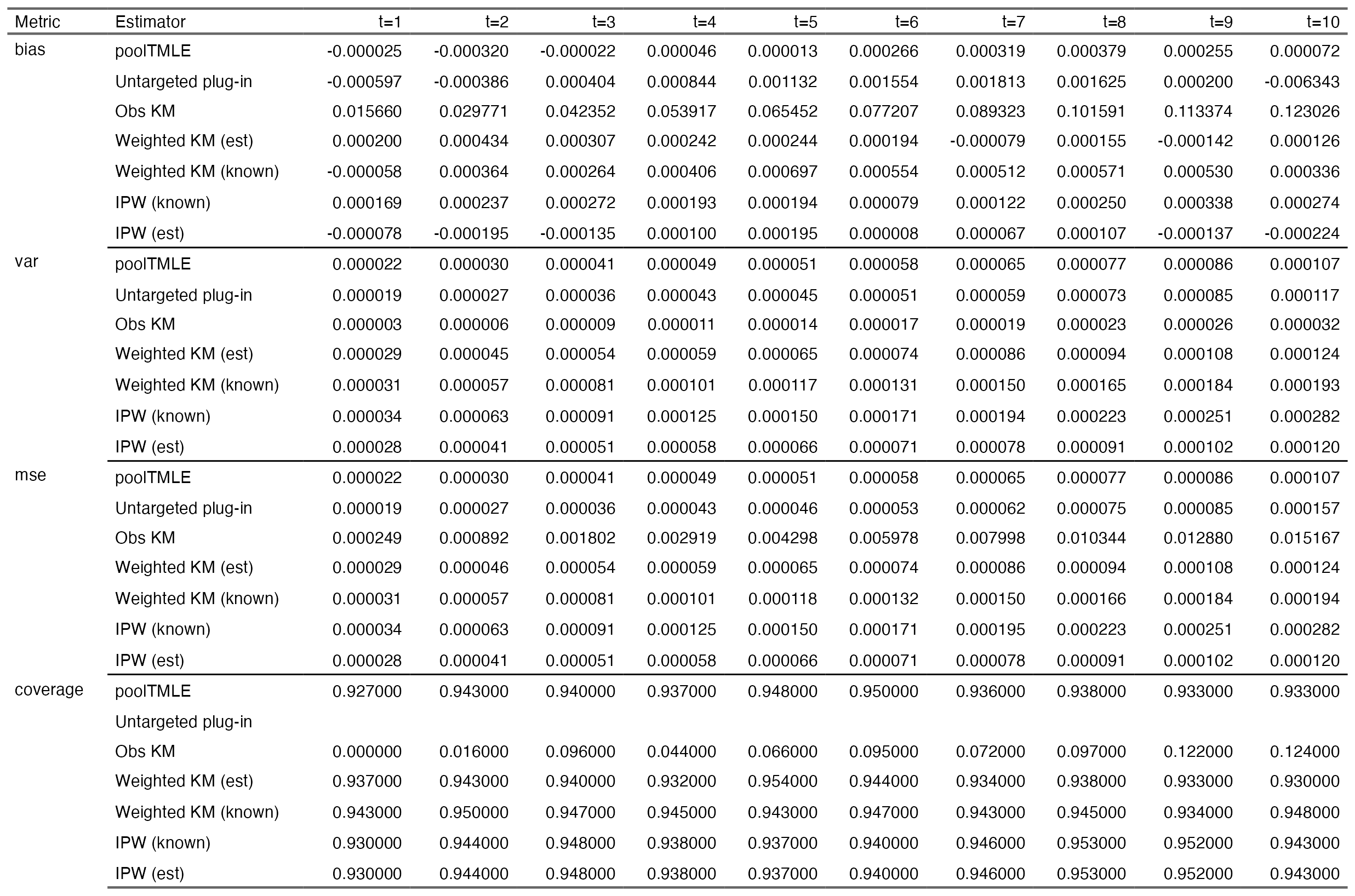}  
  \caption{Results from 1k simulation studies with sample size $N = 3$k per simulation. Variance reported is the variance of the point estimate across simulations. Inference for each estimator which supports 95\% confidence interval coverage is described in ~\ref{subsec:estimator_comp}.Here, (est) refers to estimated $\Pi_{\Delta}$ and (known) refers to the use of known values of $\Pi_{\Delta}$.}
  \label{fig:fixtau_table}
\end{sidewaysfigure}

Figure~\ref{fig:fixtau_graph} shows the variance of the point estimates for each of the ten time points across 1,000 simulations. Figure~\ref{fig:fixtau_table} provides details of the bias, variance, mean squared error, and 95\% confidence interval coverage of these estimators.

Both IPW estimators, both weighted Kaplan-Meier estimators, and the TMLE estimator are unbiased at all time points. The TMLE has the lowest variance across all time points among all estimators. The naive Kaplan-Meier estimator, which does not use resampled data, is biased as expected.

\newpage

\subsubsection{Varied Follow-up ($\tau$) Case Results}

\begin{figure}[H]
  \centering
  \includegraphics[width=\textwidth]{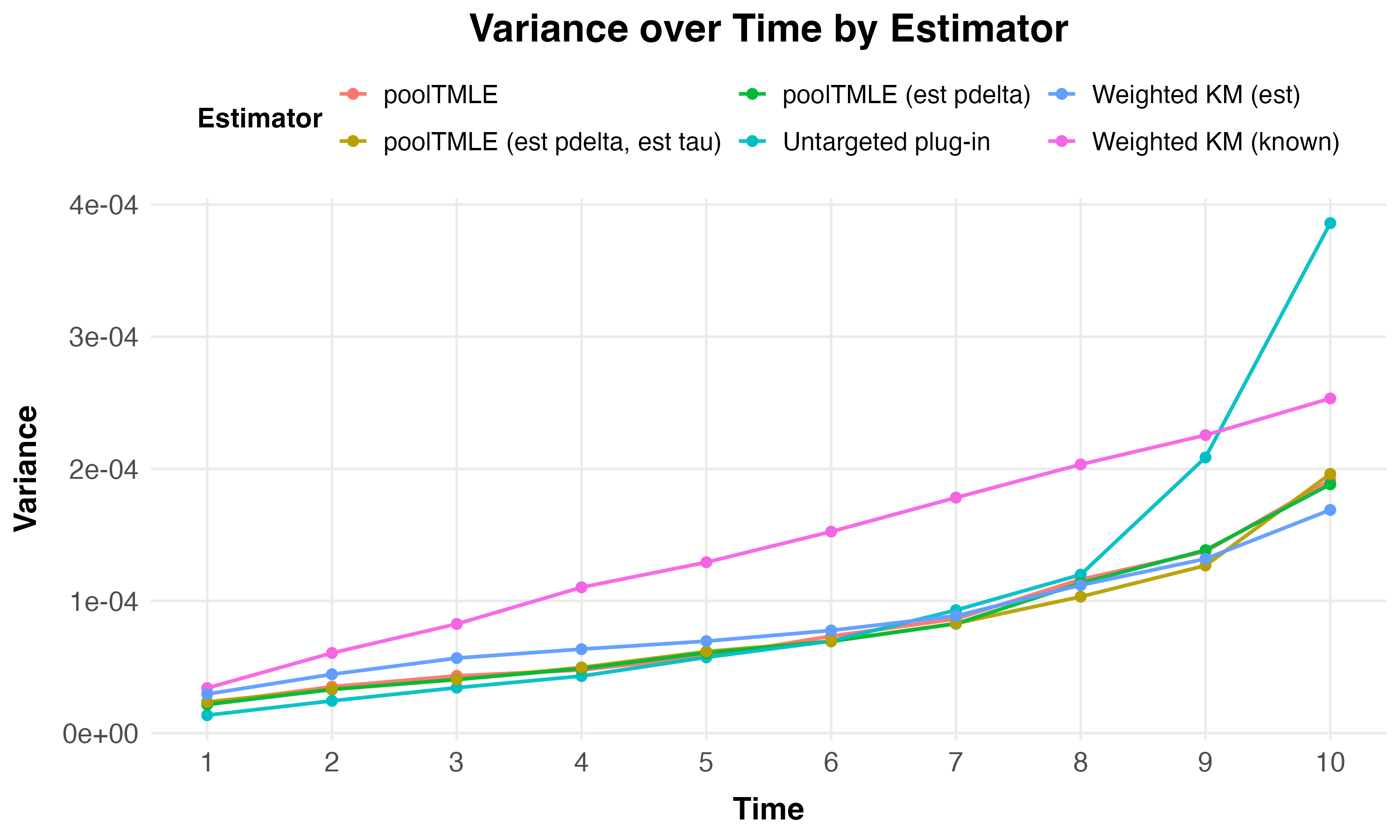}  
  \caption{Variance of unbiased estimators over time under varied $\tau$ case. Observed Kaplan-Meier omitted due to known theoretical bias in this setting. Here, (est pdelta) refers to estimated $\Pi_{\Delta}$ and (est tau) refers to estimated $\Pi_{\tau}$. (known) refers to the use of known probabilities.}
  \label{fig:randtau_graph}
\end{figure}

\begin{sidewaysfigure}
  \centering
  \includegraphics[width=\textheight]{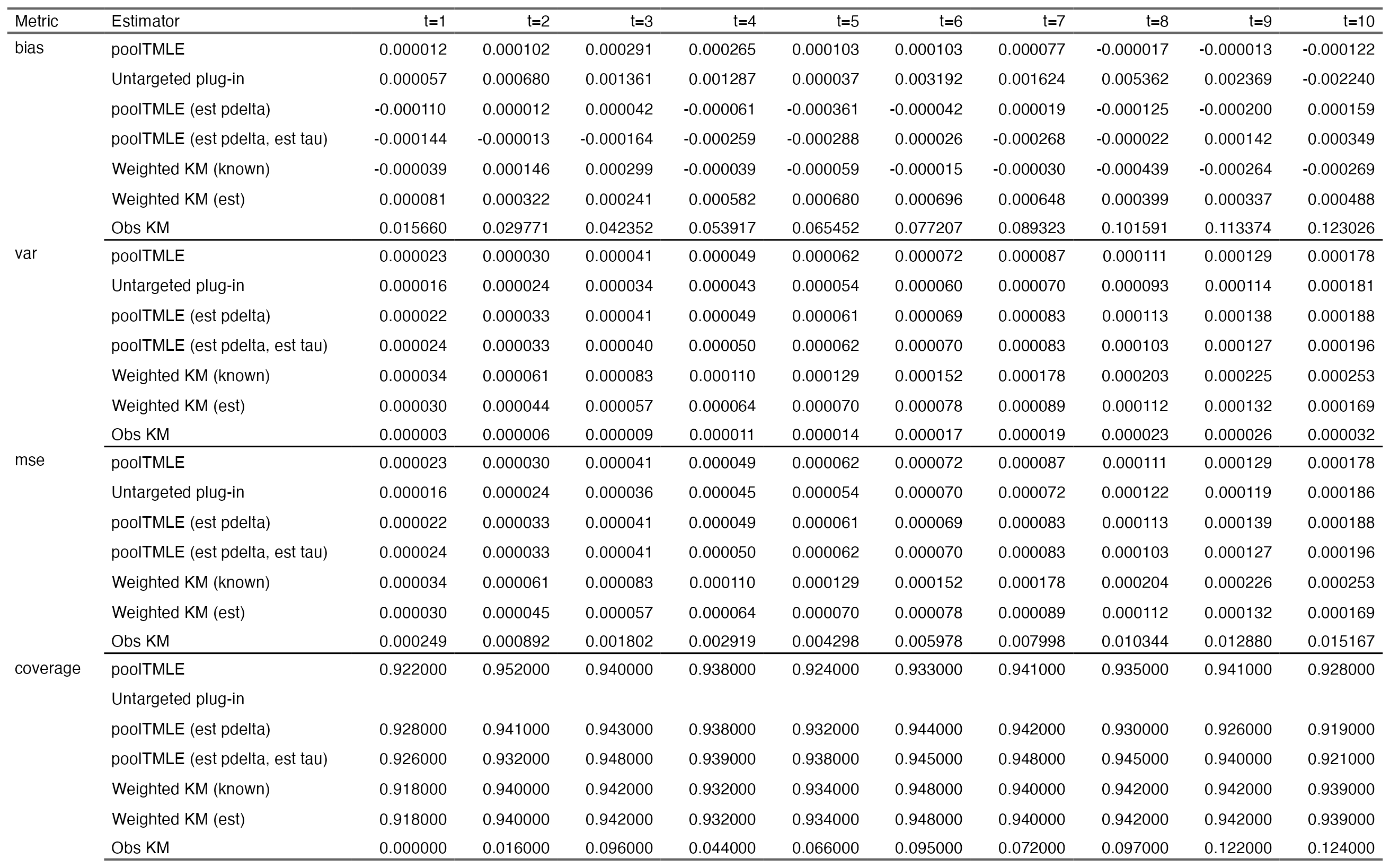}  
  \caption{Results from 1k simulation studies with sample size $N = 3$k per simulation. Variance reported is the variance of the point estimate across simulations. Inference for each estimator which supports 95\% confidence interval coverage is described in ~\ref{subsec:estimator_comp}. Here, (est pdelta) refers to estimated $\Pi_{\Delta}$ and (est tau) refers to estimated $\Pi_{\tau}$. (known) refers to the use of known probabilities.}
  \label{fig:randtau_table}
\end{sidewaysfigure}

Figure~\ref{fig:randtau_graph} shows the variance of the point estimates for ten time points for all theoretically unbiased estimators across 1,000 simulations. Figure~\ref{fig:randtau_table} provides details of the bias, variance, mean squared error, and 95\% confidence interval coverage of these estimators.

Both weighted Kaplan-Meier estimators and all TMLEs are unbiased across all time points. The naive Kaplan-Meier is again biased as expected. The variance of the TMLE with estimated probabilities for $\Delta$ and $\tau$ censoring is the lowest on average across all time points. This TMLE, in fact, has the lowest variance at all time points except for the final time points $t=10$ where the weighted Kaplan-Meier estimator with estimated weights has the lowest variance.

\newpage

\section{Discussion }

In this paper, we develop (1) a novel Targeted Maximum Likelihood Estimators (TMLE) for the survival curve in resampling designs with fixed follow-up durations and (2) an IPCW-TMLE that remains highly efficient when follow-up durations vary across individuals. This IPCW-TMLE can also be extended to handle cases where the varied follow-up times are informative and depend on measured covariates. Our simulation studies demonstrate that these TMLEs are both unbiased and substantially more efficient than existing alternatives.

In the fixed $\tau$ setting, the proposed TMLE is theoretically unbiased and efficient. Across all time points, the pooled TMLE showed markedly improved efficiency relative to commonly used weighted Kaplan--Meier and IPW estimators. On average, the pooled TMLE had approximately 2--2.5 times lower variance than estimators IPW and KM estimators which use known resampling probabilities, and 20--30\% lower variance than IPW and KM estimators that estimate these probabilities. These gains were especially pronounced early in follow-up, where our TMLE leverages the most substantial amount of future covariate information. Since weighted KM with known weights is the most commonly applied estimator in practice, these results highlight a meaningful improvement over current methodology.

In the setting with varied follow-up times, the pooled TMLE that estimates both $\Pi_{\Delta}$ and $\Pi_{\tau}$ censoring mechanisms was the most efficient estimator overall. At some time points, this TMLE reduced variance by as much as 12--13\% relative to the TMLE that uses known probabilities for both $\Pi_{\Delta}$ and $\Pi_{\tau}$, with the largest gains appearing later in follow-up when censoring is more pronounced. Compared to this best-performing TMLE, the weighted KM estimator with known weights had, on average, approximately 90\% higher variance (exceeding 120\% at multiple intermediate time points), while weighted KM with estimated censoring hazards had on average 16\% higher variance (though slightly smaller variance at $t=10$ only). These findings show that our TMLE, especially with estimation both $\Pi_{\Delta}$ and $\Pi_{\tau}$, yields substantial improvements in precision over commonly used estimators from the literature. The slightly larger variance of TMLE in this setting at $t=10$ relative to weighted KM with estimated weights is likely due to the nature of the estimation problem within TMLE, where deterministic information in combination with heavier censoring by $\tau$ at later time points reduces the effective sample size at later time points. 

The inefficiency of the weighted KM estimator is most easily illustrated by example. Suppose $t_0 = 10$ and a participant makes their last visit at $M = 9$ but is not selected for resampling. Because $\Delta = 0$, the weighted KM assigns this individual a weight of zero, effectively discarding them from the analysis, despite their visit history providing near-certain information that they remain alive at $t_0$. In contrast, our TMLE leverages the deterministic knowledge that this participant is alive through $t=9$ and simply models their hazard of death at $t=10$ based on their covariate information. This yields a substantially more efficient use of valuable, highly predictive covariate data.

In addition to our TMLE, we propose a novel extension of the weighted Kaplan-Meier estimator that replaces known resampling weights with estimated $\Pi_{\Delta}$ using a simple logistic regression model. To our knowledge, this modification has not previously been suggested in the resampling literature yet has immense potential to reduce variance, especially when there is a large sample of patients who have $\Delta=0$ and their data is thus otherwise completely ignored by the KM estimator.  

Another key insight of our TMLE approach is that the CAR likelihood structure for this problem justifies conditioning outcome hazard estimation on both past and future covariate information. As a consequence, time-dependent covariates can be treated as if they were baseline covariates in the hazard model, allowing the TMLE to leverage vast amounts of highly predictive covariate information and promosting the stability and efficiency of this approach.

While the untargeted plug-in estimator can behave reasonably well under some conditions, we recommend always including the targeting step. Without targeting, inference is not generally valid with machine learning estimators, and under known resampling probabilities TMLE enjoys a double robustness property, remaining consistent even if the initial outcome hazard model is misspecified.

Several important extensions remain. Our work focuses on a single resampling decision at study end, whereas many studies may conduct intermittent or covariate-adaptive resampling \citep{an2015choosing}. Future work could include formalizing extensions of our approach to handle these cases. Additionally, while we present a fully efficient TMLE for the fixed $\tau$ case, the TMLE we present for the random $\tau$ case is not a fully semiparametric efficient estimator, which can intuitively seen by its reliance on stratification $\tau \geq t_0$. Future work could focus on proposing a fully efficient estimator in this context. 

In summary, we provide a principled framework for efficient survival estimation under resampling designs. Our methods are grounded in rigorous semiparametric theory, demonstrate substantial efficiency gains over current approaches, and are broadly applicable to real-world studies. This work lays the foundation for constructing highly efficient estimators for survival in resampling design settings by fully leveraging all collected covariate information over time.

\newpage
\bibliography{references}
\bibliographystyle{apalike}

\clearpage
\appendix

\input{supplement}

\end{document}

%% file: supplement.tex
\section{Supplementary Methods}
\subsubsection{Tangent Space and Projection of $D^*_G$ onto $T_G(P)$} \label{subsubsec:projection}
When efficiency is gained by estimating these censoring weights for $\tau$, using $D^*_{G}(P)$ for inference for the resulting TMLE will be conservative. To derive the influence curve which provides accurate inference when the censoring mechanism is estimated, we consider the tangent space of the censoring distribution and the projection of $D^*_G$ onto this space. Here, we use $G$ to refer to the $\tau$ censoring mechanism.

Let $A(t)$ be an indicator of whether a participant is censored by time $t$, defined as $A(t) = \mathbb{I}(t > \tau)$. Let $dA(t) = \mathbb{I}(t = \tau)$ denote the increment of the censoring process at time $t$. The tangent space of the censoring distribution, denoted $T_G(P)$, can be decomposed as an orthogonal sum across time:
\[
T_G(P) = \sum_t T_{G,t}(P),
\]
where each component $T_{G,t}(P)$ corresponds to the tangent space of the conditional distribution of $A(t)$ given its past. This includes functions of $dA(t)$ and its parents, $Pa(A(t)) = (W, \bar{L}(t), \bar{A}(t-1))$, with conditional mean zero given $Pa(A(t))$.

\paragraph{Projection of $D^*_G$ onto $T_{G,t}(P)$:}
The projection of the function $D^*_G$ onto the tangent space $T_{G,t}(P)$ is given by:
\begin{align*}
    \prod(D^*_G \mid T_{G,t}(P)) &= \big[ E(D^*_G \mid dA(t) = 1, Pa(A(t))) - E(D^*_G \mid dA(t) = 0, Pa(A(t))) \big] \\
    & \quad \times \big( dA(t) - \lambda_{\tau}(t \mid W, \bar{L}(t)) \big).
\end{align*}
Let $dM_G(t)$ denote the martingale increment of the censoring process:
$$dM_G(t)=dA(t) - E[dA(t) \mid Pa(A(t))]$$
Then, the projection simplifies to:
\begin{align*}
    \prod(D^*_G \mid T_{G,t}(P)) &= \big[ E(D^*_G \mid \tau = t, Pa(A(t))) - E(D^*_G \mid \tau > t, Pa(A(t))) \big] \\
    & \quad \times dM_G(t).
\end{align*}

\paragraph{Projection of $D^*_G$ onto $T_G(P)$:}
The full projection of $D^*_G$ onto the tangent space $T_G(P)$ is the sum of projections across all time points:
\[
\prod(D^*_G \mid T_G(P)) = \sum_t \prod(D^*_G \mid T_{G,t}(P)).
\]

\begin{lemma}
This gives the influence curve for the IPCW-TMLE with estimated censoring hazard:
\begin{align*}
    D^{**}_G(t) &= D^*_G(t) - \prod(D^*_G \mid T_G(P)) \\
    &= \left[ \mathbb{E}\big(D^*_G(t) \mid \tau = t, \operatorname{Pa}(A(t))\big) 
    - \mathbb{E}\big(D^*_G(t) \mid \tau > t, \operatorname{Pa}(A(t))\big) \right] \, dM_G(t).
\end{align*}
\end{lemma}

This projection plays a role in both inference and understanding the maximum possible efficiency gains achievable by flexibly estimating or targeting the conditional hazard of censoring.

\subsection{Estimating the Projection}
\label{subsubsec:est_proj}

To implement the targeting step for the censoring mechanism $G$, we must first estimate the clever covariate $f(t)$ for each individual at each time point.  As with the outcome targeting step, this procedure can be performed in a pooled fashion, simultaneously updating the conditional hazard of censoring across all $t$ by leveraging a single regression fit.

Recall that $f(t)$ is defined as the difference in the following conditional expectations:
\[
f(t) = E(D^*_G \mid dA(t) = 1, Pa(A(t))) - E(D^*_G \mid dA(t) = 0, Pa(A(t)))
\]
To estimate these conditional expectations, we regress the estimated influence curve $D^*_G$ onto the censoring indicator $A_i(t)$ and its parents $A(t)$, $Pa(A(t)) = (W, \bar{L}(t))$ for each time point $t$:
$$D_G^*(O_i) \sim A_i(t) + \bar{L}_i(t-)$$
where $\bar{L}_i(t-)$ includes the covariate history available just prior to time $t$. Though, one could alternatively choose to agree on common summary measures of the past and fit this regression in a pooled mannner. One may choose any estimator (from a linear model through flexible ensemble machine learning via Super Learner). 

Once this model is fit, it is evaluated at both levels of the censoring indicator, $A_i(t) = 1$ and $A_i(t) = 0$, to obtain the estimated values of $f(t)$ to be used in the targeting step. 

\subsection{Joint Targeting Procedure: Conditional hazard of death and conditional hazard of censoring} \label{subsubsec:joint_targeting}
When implementing targeting for both the conditional hazard of death and the conditional hazard of censoring, these targeting steps must iterate back and forth. More concretely, the steps of this joint targeting procedure are as follows:
\begin{enumerate}
    \item Get initial estimates of the conditional hazard of censoring $\lambda_{\tau}(t \mid W, \bar{L}(t))$ using SuperLearner
    \item Get initial estimates of the conditional hazard of death $\lambda_T(t \mid W, \bar{L}(\tau))$ using SuperLearner
    \item Perform a targeting step for the conditional hazard of death $\lambda_T(t \mid W, \bar{L}(\tau))$. Compute $D_G^*$ using the targeted values $\lambda_T^*(t \mid W, \bar{L}(\tau))$.
    \item Use the estimated $D_G^*$ to estimate the clever covariate $f(t)$.
    \item Perform a targeting step for the conditional hazard of censoring $\lambda_{\tau}(t \mid W, \bar{L}(t))$ using the estimated $f(t)$. 
    \item Repeated steps 3-5 until both targeting steps reach their respective convergence criteria.
    \item Run one last targeting step for the hazard of death $\lambda_T(t \mid W, \bar{L}(\tau))$ using the final values for the targeted conditional hazard of censoring $\lambda_{\tau}(t \mid W, \bar{L}(t))$ and get the final value of $D_G^*$ to be used for inference.
\end{enumerate}

\subsection{Simulation Setup}
\label{subsubsec:sim_setup}

We simulate a cohort of $n$ individuals with discrete follow-up at time points $t = 1, \ldots, 10$. Three baseline binary covariates $W_1, W_2, W_3 \sim \text{Bernoulli}(0.5)$ represent risk and protective factors.

\paragraph{Clinic visit process.}
At each time $t$, individuals may attend a clinic visit according to
\[
V_t \sim \text{Bernoulli}\!\left( 
\text{logit}^{-1}\!\left(
\begin{aligned}
&1.0\,W_1 + 1.0\,W_2 - 1.0\,W_3 
+ \mathbf{V}_{t-1}^\top \boldsymbol{\alpha} \\
&\quad - 0.05\,I(L_{t-1}^{\mathrm{under}} < 200)
- 0.05\,I(L_{t-1}^{\mathrm{under}} < 100)
\end{aligned}
\right)\right),
\]
where $\mathbf{V}_{t-1} = (V_{t-1}, V_{t-2}, V_{t-3})$ and $\boldsymbol{\alpha} = (0.4,0.3,0.2)$. Individuals cannot attend visits after death, so $V_t = 0$ if $D_t = 1$.

\paragraph{Continuous-valued biomarker process.}
An underlying CD4-like marker evolves via an autoregressive process:
\[
L_t^{\mathrm{under}}
= 0.8\,L_{t-1}^{\mathrm{under}} + 0.2\,\mu_t + \varepsilon_t,
\qquad \varepsilon_t \sim N(0, 15^2),
\]
where
\[
\mu_t =
\begin{aligned}[t]
&200 - 100\,W_1 + 100\,W_2 - 100\,W_3 \\
&+ (10,15,10)\cdot (V_t, V_{t-1}, V_{t-2}) \\
&- 5\,I(L_{t-1}^{\mathrm{under}} < 200)
- 10\,I(L_{t-1}^{\mathrm{under}} < 100).
\end{aligned}
\]
Values are truncated to $[20,1500]$. Observed CD4 counts are
\[
L_t = L_t^{\mathrm{under}} \cdot I(V_t = 1).
\]

\paragraph{Death process.}
Death occurs according to
\[
D_t \sim \text{Bernoulli}\!\left( 
\text{logit}^{-1}\!\left(
\begin{aligned}
&-4.5 + 0.065(t-1)
+ 1.0\,W_1 - 1.0\,W_2 + 1.0\,W_3 \\
&+ \mathbf{V}_{t-1}^\top \boldsymbol{\gamma}
+ 0.1\,I(L_{t-1}^{\mathrm{under}} < 200)
+ 0.3\,I(L_{t-1}^{\mathrm{under}} < 100)
\end{aligned}
\right)\right),
\]
where $\boldsymbol{\gamma} = (-0.3,-0.2,-0.2)$. Once dead, always dead: $D_t = 1$ if $D_{t-1} = 1$.

\paragraph{Death identification.}
For individuals who newly die at time $t$,
\[
I(D_t = 1 \;\&\; D_{t-1} = 0) = 1,
\qquad 
\text{Id}_t \sim \text{Bernoulli}(0.2),
\]
and $\text{Id}_t = 1$ persists for all future times once observed.

\paragraph{Random follow-up horizon.}
Each individual has a limited follow-up time
\[
\tau \in \{5,7,9,10\}
\]
with probabilities $(0.10, 0.15, 0.15, 0.60)$. If under the fixed $\tau$ case, then $\tau=10$ for all participants.

\paragraph{Censoring rules.}
All time-varying variables are deterministically censored beyond $\tau$:
\[
D_t = 0,\quad V_t = 0,\quad L_t = 0,\quad \text{Id}_t = 0
\qquad \text{for } t > \tau.
\]
Natural death observed by $\tau$ is recorded as:
\[
\text{nat\_obs\_death} = I(\exists\, t \le \tau : \text{Id}_t = 1).
\]

\paragraph{Outcome observability.}
Let $M = \max\{t : V_t = 1\}$ be the last observed visit time.
\[
\Delta = I(M \ge \tau \;\text{or}\; \text{nat\_obs\_death} = 1).
\]
If $M < \tau$ and natural death is not observed, death may still be observed via administrative resampling with probability $0.2$. Second-stage outcomes are set to NA when $\Delta = 0$ even after resampling is complete.

\paragraph{Observed data.}
Thus the observed data structure is
\[
O = 
\big(
W_1, W_2, W_3,
\{D_t, V_t, L_t, \text{Id}_t\}_{t=1}^{10},
\tau, \Delta, \text{second-stage outcomes}
\big).
\]